Thermoregulation in mice, rats and humans: An insight into the evolution of human hairlessness

Bernard J. Feldman

Department of Physics and Astronomy

University of Missouri-St. Louis, St. Louis, MO USA

Abstract

The thermoregulation system in mammals removes body heat in hot temperatures and retains body heat in cold temperatures. The better the animal removes heat, the worse the animal retains heat and visa versa. It is the balance between these two conflicting goals that determines the mammal's size, heart rate and amount of hair. The rat's loss of tail hair and human's loss of its body hair are responses to these conflicting thermoregulation needs as these animals evolved to larger size over time.

There have been numerous ideas about why humans lost most of their hair. They include increased dissipation of body heat for living on the open savannah [1] response to a evolving , metabolically active brain [2], improved heat dissipation needed for the transition to bipedalism, leading to faster running and better success at hunting [3], faster swimming by an aquatic ape [4], sexual selection, i.e., female preference of males with less hair [5], use of clothing [6], reduction of parasites (lice, fleas and ticks) that live in the hair [7], discovery and use of fire [8], neotony—the tendency of humans to retain infant characteristics, e.g., less hair, for longer periods [9], enhanced social communication [10], parental selection [11], allometry—the study of body size and physiology combined with the observation that larger apes have less dense hair [12], and pleasure of skin to skin contact between mother, partner and child. [13]

This manuscript proposes that a combination of several of these ideas—improved heat removal, allometry, improved success at hunting and sexual preferences—can explain why humans became largely hairless. At the heart of this manuscript is a discussion of the regulation of body temperature. All mammals need to regulate their body temperatures. They need to remove heat in hot environments and retain heat in cold environments. This temperature regulation is controlled primarily by the circulatory system, consisting of blood flowing through arteries, veins, the heart and the lungs. In hot climates, blood carries heat to the skin and the lungs, where heat escapes to the environment by conduction and evaporation to the air next to the skin and inside the lungs. In cold climates, warm blood is carried from the inner, warmer parts of the body to the colder outer parts of the body.

The vast majority of land-based mammals have fur covering almost all of their skin. This keeps the animal warm in cold temperatures at the expense of cooling in hot temperatures. This strategy makes sense for two reasons: First, other strategies keep an individual cool in the heat—take a dip in a lake, sit in the shade or take a nap. Second, because fur retains body heat, it dramatically reduces the food the individual needs to gather or hunt. In contrast, few strategies are available to keep warm in cold temperatures—live in a cave or underground. For *Homo habilis*, the evidence suggests that hair loss began more than 1.2 million years ago, even though the last ice age started about 1.6 million years ago.[7] So *Homo habilis* was losing hair even as individuals had to deal with ever cooling temperatures and increasingly cold evenings. Evidence suggests that the use of fire began about 0.5 million years ago and clothing about 0.3 million years ago, well after major climate cooling.[8]

So why would any land based mammal give up the advantages of having fur? Let us consider the examples of rats and mice. Rats and mice differ in three important ways that are relevant to this discussion. Rats are much larger than mice (84 day old rats weigh 183-409 gm; 84 day old mice, 18 - 32 gm).[14] Rats have a slower heart beat (330 - 480/minute) than mice (310 – 840/minute).[15] Rats have a hairless tail from which they dissipate about 17% of their body heat while mice tails are covered in fur.[16]

Simply put, the smaller the animal, the higher the ratio of surface area of the skin and lungs to body volume and the shorter the distance from hot blood vessels to the cold air next to the skin or inside the lungs. Smaller animals lose heat more rapidly, requiring more fur to insulate the animal and a higher heart rate to generate heat from the beating heart muscles and to circulate warm blood more rapidly to the animal's cold exterior. In contrast, for larger animals, heat retention is less of a problem and heat removal becomes more of a problem. The evolution of a hairless tail in the rat is one solution to the problem of heat removal.

Now let us consider the largely hairless human. It is estimated that the average adult height of *Homo habilis* was no more than 1.3 m (4 feet 3 inches) tall and the brain volume of 600 cm$^3$.[2] Modern man is about 1.7 m (5 feet, 6 inches) tall, on average, and has a brain volume of 1300 cm$^3$. [17] The brain generates a great deal of heat and the larger the brain, the greater the problem of heat removal.[2] As human ancestors got larger in both overall size and brain size, the problem of heat removal became more important than heat retention. So just like rats' tails, human skin started to lose its hair with a concomitant increase in heat loss.

However, as *Homo habilis* got larger and lost hair, it needed more food to keep warm during periods of inactivity. Improved heat dissipation of *Homo habilis* permitted hunting in the heat of day, running faster and chasing game for longer periods. Also, the larger brain led to better hunting strategies and techniques .

Finally, the driving force behind the ever increasing size of hominoids is probably sexual selection--larger males became alpha males and had greater access to females, thus creating evolutionary pressure for increasing body size. And over a million years, larger body size confronted hominoids with an increasing problem of heat removal and the evolutionary response of reduced body hair.

Clearly, these ideas are very speculative. So are all the other hypotheses addressing the loss of hair by our distant ancestors. Also, many mammals evolved larger body size without losing their hair. However, the one thing this theory has going for it is the hairless rat tail—an example of another mammal that lost its hair to improve heat removal as it evolved to a larger size.

I would like to acknowledge the invaluable assistance of Robert Ricklefs in the preparation of this manuscript.

References

[1]. P. Wheeler, "The influence of the loss of functional body hair on hominoid energy and water budgets," Journal of Human Evolution 223, 379-388 (1992).

[2]. N. Jablonski , https://www.independent.co.uk/news/science/why-humans-lost-their-body-hair-to-stop-their-brains-from-overheating-as-we-evolved-8498623.html .

[3]. P. Wheeler, "The evolution of bipedality and loss of functional body hair in hominoids, Journal of Human Ecolution 13, 91-98 (1984).

[4]. A. Hardy, "Was man more aquatic in the past?" New Scientist 7, 642-645 (1960) and E. Morgan, "The aquatic ape hypothesis," Souvenir Press (London) 1997.

[5]. C. Darwin, "The descent of man, and selection in relation to sex," John Murray (London) 1871.

[6]. J. A. Kushlan, "The vestiary hypothesis of human hair reduction," Journal of Human Evolution 14, 29-32 (1985).

[7]. M. Pagel and W. Bodmer, https://www.nytimes.com/2003/08/19/science/why-humans-and-their-fur-parted-ways.html

[8]. A. J. Couch, https://peerj.com/preprints/1702.pdf.

[9]. Steven J. Gould, "Ontogeny and Phylogeny," Harvard University Press (Cambridge), 1977.

[10]. N. Simmons, "What is the difference between hair and fur? Scientific American (2001), https://www.scientificamerican.com/article/what-is-the-difference-be/.

[11]. J. R. Harris, "Parental selection: a third selection process in the evolution of human hairlessness and skin color," Medical Hypotheses 66, 1053-1059 (2006).

[12]. G. G. Schwartz and R. A. Rosenblum, "Allometry of primate hair density and the evolution of human hairlessness," American Journal of Physical Anthropology 55, 9-12 (1981).

[13]. J. Giles, Naked love: the evolution of human hairlessness, Biological Theory 5, 326-336 (2011).

[14]. http://www.arc.wa.gov.au/?page_id=125.

[15]. https://www.google.com/search?source=hp&ei=JzzrW9fJO6bojgSwvK-oCQ&q=Heart+rate+of+mice+and+rats&btnK=Google+Search&oq=Heart+rate+of+mice+and+rats&gs_l=psy-ab.3..33i22i29i30.1581.6621..7124...0.0..0.126.2434.19j8......0....1..gws-wiz.....0..0j0i131j0i22i30j0i22i10i30.9CC2vzGIACU.

[16]. http://www.ratbehavior.org/RatTails.htm.

[17]. https://en.wikipedia.org/wiki/Homo_habilis